\documentclass[10pt,final,letterpaper]{IEEEtran}
\usepackage{amsmath}
\usepackage{latexsym}
\usepackage{graphicx}
\usepackage{bbding}
\usepackage{indentfirst}
\usepackage{cases}
\usepackage{algorithm2e}
\usepackage{subeqnarray}
\IEEEoverridecommandlockouts

\begin{document}
\title{Energy-Efficient Optimization for Wireless Information and Power Transfer in Large-Scale MIMO Systems Employing Energy Beamforming}
\author{\authorblockN{Xiaoming~Chen, Xiumin~Wang and Xianfu~Chen
\thanks{This work was supported by the Natural Science Foundation of
China (No. 61301102, 61300212), the Natural Science Foundation of
Jiangsu Province (No. BK20130820), the open research fund of
National Mobile Communications Research Laboratory, Southeast
University (No. 2012D16) and the Doctoral Fund of Ministry of
Education of China (No. 20123218120022).}
\thanks{Xiaoming~Chen (e-mail: {\tt chenxiaoming@nuaa.edu.cn})
is with the College of Electronic and Information Engineering,
Nanjing University of Aeronautics and Astronautics, and also with
the National Mobile Communications Research Laboratory, Southeast
University, China. Xiumin~Wang (e-mail: {\tt wxiumin@hfut.edu.cn})
is with the School of Computer and Information, Hefei Univesity of
Technology, Hefei, China. Xianfu~Chen (e-mail: {\tt
xianfu.chen@vtt.fi}) is with the VTT Technical Research Centre of
Finland, Oulu, Finland.}}}

\markboth{IEEE Wireless Communications Letters, Vol. XX, No. Y,
Month 2013}{} \maketitle

\begin{abstract}
In this letter, we consider a large-scale multiple-input
multiple-output (MIMO) system where the receiver should harvest
energy from the transmitter by wireless power transfer to support
its wireless information transmission. The energy beamforming in the
large-scale MIMO system is utilized to address the challenging
problem of long-distance wireless power transfer. Furthermore,
considering the limitation of the power in such a system, this
letter focuses on the maximization of the energy efficiency of
information transmission (bit per Joule) while satisfying the
quality-of-service (QoS) requirement, i.e. delay constraint, by
jointly optimizing transfer duration and transmit power. By solving
the optimization problem, we derive an energy-efficient resource
allocation scheme. Numerical results validate the effectiveness of
the proposed scheme.
\end{abstract}

\begin{keywords}
Energy-efficient resource allocation, wireless information and power
transfer, large-scale MIMO system, energy beamforming, QoS
guarantee.
\end{keywords}

\section{Introduction}
Wireless power transfer facilitates the battery charging to prolong
the lifetime of wireless networks, especially under some extreme
conditions, such as battlefield, underwater and body area networks
\cite{Medical}. Thus, wireless power transfer draws considerable
research attentions from academics and industry
\cite{WET0}-\cite{WET3}.

In general, wireless power transfer from a power source to a
receiver is implemented through electromagnetic propagation
\cite{WET4}. Similar to wireless information transmission, wireless
power transfer also suffers from propagation loss, including path
loss, shadowing and fast fading. Therefore, transfer efficiency is a
critical and challenging issue for wireless power transfer. To solve
this problem, multi-antenna techniques are introduced into wireless
power transfer to improve the transfer efficiency by using energy
beamforming. In \cite{EnergyBeamforming1}, the design methods of the
optimal transmit beam for power transfer were given in MIMO
broadcast systems. Furthermore, considering the imperfect channel
state information at the power source, a robust energy beamforming
strategy was proposed in \cite{EnergyBeamforming2}. However, due to
the limited number of antennas at the power source, the energy
efficiency based on the traditional multi-antenna systems is
difficult to satisfy the practical requirement, especially when the
transfer distance is not so short \cite{EnergyEfficiency}. Recently,
large-scale MIMO deploying a huge number of antenna at the
transmitter is proposed to enormously improve the transmission
performance by exploiting its large array gain
\cite{LargescaleMIMO1} \cite{LargescaleMIMO2}. For example, in the
60GHz WiFi 802.11/ac, since small size antenna is possible,
large-scale antenna beamforming is exactly what is being
implemented. Similarly, large-scale MIMO can be used to effectively
improve the performance of wireless power transfer.

Intuitively, the ultimate purpose of wireless power transfer is to
fulfill the need of the receiver for work. As a simple example, in
medical field, the implanted equipment in body is powered through
wireless power transfer, and then transmits the medical data to the
outside receiver with the harvested energy. However, most of
previous analogous works analyze and design the wireless power
transfer without taking into consideration the utilization of the
harvested energy. In this letter, we consider joint wireless
information and power transfer in large-scale MIMO systems employing
energy beamforming, where the power receiver transmits the
information with the harvested energy, namely wireless powered
communication. Since energy harvesting and information transmission
are impossibly carried out simultaneously, time slot should be
divided into the harvesting and the transmitting components. In
order to optimize the performance, it is necessary to determine the
optimal time switch point, namely time resource allocation.
Moreover, transmit power at the power source, as another important
resource, also affects the ultimate performance. Recently,
energy-efficient communications, namely green communication, gain
more and more prominence due to energy shortage and greenhouse
effect \cite{Green}. This letter focuses on the maximization of
energy efficiency, defined as the delivered information bits per
Joule harvested energy, by jointly optimizing transfer duration and
transmit power. The contributions of this letter lie in three folds.
First, it solves the challenging problem of long-distance wireless
information and power transfer. Second, it improves the energy
efficiency enormously. Third, it provides high QoS guarantee.

The rest of this letter is organized as follows. We first give an
overview of wireless information and power transfer in large-scale
MIMO systems in Section II, and then derive an energy-efficient
resource allocation scheme by maximizing the energy efficiency while
satisfying the QoS requirement in Section III. In Section IV, we
present some numerical results to validate the effectiveness of the
proposed scheme. Finally, we conclude the whole paper in Section V.

\section{System Model}
\begin{figure}[h] \centering
\includegraphics [width=0.4\textwidth] {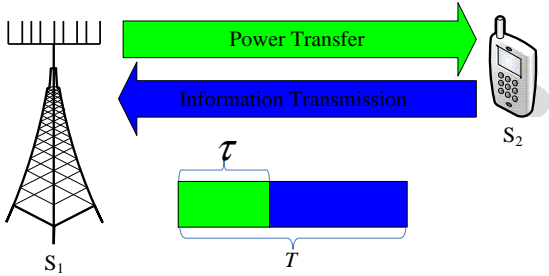}
\caption {An overview of the wireless information and power system.}
\label{Fig1}
\end{figure}

We consider a time division duplex (TDD) large-scale MIMO system
employing wireless information and power transfer, as shown in
Fig.\ref{Fig1}, where the power source and the information receiver,
namely S$_1$, is equipped with $N_t$ antennas to enhance the power
transfer efficiency by energy beamforming and improve the
information transmission rate by receive combining, respectively.
Note that the number of antennas $N_t$ is quite large in the
large-scale MIMO system. Similarly, S$_2$ plays the roles as both
power receiver and information transmitter. Due to space limitation,
S$_2$ deploys a single antenna for energy harvesting and information
transmitting. S$_2$ only has limited power to maintain the active
state, so it needs to harvest enough energy from the outside for
information transmission, such as S$_1$. Considering the limited
storage, S$_2$ should be charged from S$_1$ slot by slot.

The whole system is operated in slotted time of length $T$. At the
beginning of each time slot, S$_1$ performs wireless power transfer
by using energy beamforming with the duration $\tau$. With the
harvesting energy, S$_2$ transmits the information to S$_1$ in the
rest of the time slot $T-\tau$. We use $\sqrt{\alpha}\textbf{h}$ to
denote the $N_t$ dimensional channel vector from S$_1$ to S$_2$,
where $\alpha$ represents the distance-dependent path loss and
$\textbf{h}$ is the channel fast fading vector with independent and
identically distributed (i.i.d.) zero mean and unit variance
circular symmetric complex Gaussian random entries. According to the
law of energy conservation, the harvested power at S$_2$ from S$_1$
can be expressed as \cite{EnergyBeamforming1}
\begin{equation}
P_{harv}=\eta\alpha P|\textbf{h}^H\textbf{w}|^2\label{eqn1}
\end{equation}
where $P$ is the transmit power of S$_1$, $\eta$ is the efficiency
ratio at S$_2$ for converting the harvested energy to the electrical
energy to be stored, which depends on the efficiency of the energy
converter. Intuitively, $\eta$ belong to the interval $[0,1]$.
$\textbf{w}$, as the energy beamforming vector with unit norm, is
used to adaptively adjust the energy transmit direction according to
the instantaneous CSI $\textbf{h}$, so as to maximize $P_{harv}$. We
assume S$_1$ has full CSI $\textbf{h}$ by estimating the uplink
channel from S$_2$ to S$_1$ in the last time slot, due to channel
reciprocity in TDD systems. Clearly,
$\textbf{w}=\textbf{h}/\|\textbf{h}\|$, namely maximum ratio
transmission (MRT), can maximize $P_{harv}$. Thus, the total
harvested energy at S$_2$ during a time slot is $Q_{harv}=\eta\alpha
P\|\textbf{h}\|^2\tau$. After that, S$_1$ and S$_2$ toggle to the
wireless information transfer mode. With the harvesting energy
$Q_{harv}$, S$_2$ transmits the information to S$_1$ in the rest of
the time slot $T-\tau$, and the receive signal can be expressed as
\begin{equation}
\textbf{y}=\sqrt{\frac{Q_{harv}}{T-\tau}\theta}\textbf{g}s+\textbf{n}\label{eqn2}
\end{equation}
where $s$ is the normalized Gaussian distributed transmit signal,
$\textbf{y}$ is the $N_t$ dimensional receive signal vector, and
$\textbf{n}$ is the additive Gaussian white noise with zero mean and
variance matrix $\sigma^2\textbf{I}_{N_t}$.
$\sqrt{\theta}\textbf{g}$ denotes the unlink channel from S$_2$ to
S$_1$, where $\theta$ is the distance-dependent path loss, which may
be the same as $\alpha$, and $\textbf{g}$ is the channel fast fading
vector distributed as $\mathcal{CN}(0,\textbf{I}_{N_t})$. Note that
$\textbf{g}$ may be the same as $\textbf{h}$ during a time slot due
to channel reciprocity in TDD system.
$\sqrt{\frac{Q_{harv}}{T-\tau}}$ is the transmit power. Assuming
perfect CSI at S$_1$, maximum ratio combining (MRC) is performed to
maximize the information transmission rate. Assuming there is always
data to be transmitted in each slot, then the information
transmission rate is given by
\begin{eqnarray}
R(P,\tau)&=&W\log_2\left(1+\frac{Q_{harv}\theta\|\textbf{g}\|^2}{(T-\tau)\sigma^2}\right)\nonumber\\
&=&W\log_2\left(1+\frac{\eta\alpha\theta
P\tau\|\textbf{h}\|^2\|\textbf{g}\|^2}{(T-\tau)\sigma^2}\right)\label{eqn3}
\end{eqnarray}
where $W$ is the spectrum bandwidth. As mentioned earlier, in the
large-scale MIMO system, the BS is equipped with a large number of
antennas. According to the fact that
$\lim\limits_{N_t\rightarrow\infty}\frac{\|\textbf{g}\|^2}{N_t}=1$
and
$\lim\limits_{N_t\rightarrow\infty}\frac{\|\textbf{h}\|^2}{N_t}=1$,
namely channel hardening \cite{ChannelHardening}, we have
\begin{eqnarray}
R(P,\tau)&=&W\log_2\left(1+\frac{\eta\alpha\theta N_t^2
P\tau}{(T-\tau)\sigma^2}\right)\nonumber\\
&\approx&W\log_2\frac{\eta\alpha\theta N_t^2
P\tau}{(T-\tau)\sigma^2}\label{eqn4}
\end{eqnarray}
where (\ref{eqn4}) holds true since the constant 1 becomes
negligible when $N_t\rightarrow\infty$. Because only duration
$T-\tau$ is used for information transmission during a time slot,
the average information transmission rate can be expressed as
\begin{equation}
\bar{R}(P,\tau)=\frac{T-\tau}{T}W\log_2\frac{\eta\alpha\theta
N_t^2 P\tau}{(T-\tau)\sigma^2}\label{eqn5}
\end{equation}

\emph{Remark}: The performance of wireless information and power
transfer in traditional MIMO system is limited by the path loss
$\alpha$ and $\theta$, and thus it is only applicable for
short-distance wireless information and power transfer. However, in
the large-scale MIMO system, this challenging problem can be solved
by adding antennas at the transmitter, so it can support relative
long-distance transmission and high QoS guaranteed wireless services
with a low transmit power.

\section{Energy-Efficient Power Allocation}
Considering the limitation of power resource and the requirement of
green communications, energy efficiency becomes an important
performance metric in wireless communications, especially for
wireless power transfer systems. In this section, we attempt to
derive a joint power and time resources allocation scheme to
maximize the energy efficiency while satisfying power, time and QoS
constraints, which is equivalent to the following optimization
problem:
\begin{eqnarray}
J_1: \max&& \frac{\bar{R}(P,\tau)T}{P_0T+P\tau}\label{eqn6}\\
\textmd{s.t.}&& P\leq P_{1,\max}\label{eqn7}\\
&&\frac{Q_{harv}}{T-\tau}\leq P_{2,\max}\label{eqn8}\\
&&\tau\leq T\label{eqn9}\\
&&\bar{R}(P,\tau)\geq r_{\min}\label{eqn10}
\end{eqnarray}
where $P$ is the power consumption in the power amplifier at S$_1$,
and $P_0$ is the constant power consumption in the transmit filter,
mixer, frequency synthesizer, and digital-to-analog converter which
are independent of the actual transmit power. (\ref{eqn6}) is the so
called energy efficiency, defined as information transmission bits
per Joule. (\ref{eqn7}) is the transmit power constraint at S$_1$,
(\ref{eqn8}) is the transmit power constraint at S$_2$, (\ref{eqn9})
is the time constraint at S$_1$, and (\ref{eqn10}) is the QoS
constraint, where $r_{\min}$ is the minimum transmission rate
meeting a given QoS requirement, such as delay provisioning
\cite{QoS}. Due to channel hardening in large-scale MIMO systems,
the constraint condition (\ref{eqn8}) is equivalent to
$P\leq\frac{P_{2,\max}(T-\tau)}{\eta\alpha N_t\tau}$. According to
(\ref{eqn7}), $P$ must be less than or equal to $P_{1,\max}$, so the
following condition should be satisfied
$\frac{P_{2,\max}(T-\tau)}{\eta\alpha N_t\tau}\geq P_{1,\max}$.
Thus, we have $\tau\leq\frac{P_{2,\max}T}{\eta\alpha
N_tP_{1,\max}+P_{2,\max}}=\tau_{\max}$.

The objective function (\ref{eqn6}) in a fractional program is a
ratio of two functions of the optimization variables $P$ and
$\tau$, resulting in $J_1$ is a fractional programming problem,
which is in general nonconvex. Following
\cite{EnergyEfficientPLS}, the objective function is equivalent to
$\bar{R}(P,\tau)T-q^{\star}(P_0T+P\tau)$ by exploiting the
properties of fractional programming, where $q^{\star}$ is the
energy efficiency when $P$ and $\tau$ are equal to the optimal
value $P^{\star}$ and $\tau^{\star}$ of $J_1$ respectively, namely
$q^{\star}=\frac{\bar{R}(P^{\star},\tau^{\star})T}{P_0T+P^{\star}\tau^{\star}}$.
Thus, $J_1$ is transformed as
\begin{eqnarray}
J_2: \max&& \bar{R}(P,\tau)T-q^{\star}(P_0T+P\tau)\label{eqn11}\\
\textmd{s.t.}&& P\leq P_{1,\max}\label{eqn12}\\
&&\tau\leq\tau_{\max}\label{eqn22}\\
&&\tau\leq T\label{eqn13}\\
&&\bar{R}(P,\tau)\geq r_{\min}\label{eqn14}
\end{eqnarray}
It can be proved that
$\frac{\partial^2\bar{R}(P,\tau)}{\partial^2P}<0$ and
$\frac{\partial^2\bar{R}(P,\tau)}{\partial^2\tau}<0$, so $J_2$ is a
convex optimization problem, which can be solved by the Lagrange
multiplier method. By some arrangements, its Lagrange dual function
can be written as
\begin{eqnarray}
\mathcal{L}(\mu,\nu,\upsilon,P,\tau)&=&\bar{R}(P,\tau)T-q^{\star}(P_0T+P\tau)-\mu
P\nonumber\\
&+&\mu P_{1,\max}-\vartheta\tau+\vartheta\tau_{\max}-\nu\tau+\nu
T\nonumber\\ &+&\upsilon\bar{R}(P,\tau)-\upsilon
r_{\min}\label{eqn15}
\end{eqnarray}
where $\mu\geq0$, $\vartheta\geq0$, $\nu\geq0$ and $\upsilon\geq0$
are the Lagrange multipliers corresponding to the constraint
(\ref{eqn12}), (\ref{eqn22}), (\ref{eqn13}) and (\ref{eqn14}),
respectively. Therefore, the dual problem of $J_2$ is given by
\begin{eqnarray}
\min\limits_{\mu,\vartheta,\nu,\upsilon}\max\limits_{P,\tau}\mathcal{L}(\mu,\vartheta,\nu,\upsilon,P,\tau)\label{eqn16}
\end{eqnarray}
Given $\mu$, $\vartheta$, $\nu$ and $\upsilon$, the optimal transmit
power $P^{\star}$ and transfer duration $\tau^{\star}$ can be
derived by jointly solving the following KKT conditions
\begin{eqnarray}
\frac{\partial\mathcal{L}(\mu,\vartheta,\nu,\upsilon,P,\tau)}{\partial
P}&=&(T+\upsilon)\frac{\partial\bar{R}(P,\tau)}{\partial
P}-q^{\star}\tau-\mu\nonumber\\
&=&\frac{W(T+\upsilon)(T-\tau)}{\ln2TP}-q^{\star}\tau-\mu\nonumber\\
&=&0\label{eqn17}
\end{eqnarray}
and
\begin{eqnarray}
\frac{\partial\mathcal{L}(\mu,\vartheta,\nu,\upsilon,P,\tau)}{\partial\tau}&=&(T+\upsilon)\frac{\partial\bar{R}(P,\tau)}{\partial
\tau}-q^{\star}P-\vartheta-\nu\nonumber\\
&=&W(T+\upsilon)\bigg(\frac{1}{(T-\tau)\ln2}\nonumber\\
&-&\frac{1}{T}\log_2\frac{\eta\alpha\theta
N_t^2P\tau}{(T-\tau)\sigma^2}\bigg)-q^{\star}P-\vartheta-\nu\nonumber\\
&=&0\label{eqn18}
\end{eqnarray}
Moreover, $\mu$, $\vartheta$, $\nu$ and $\upsilon$ can be updated by
the gradient method, which are given by
\begin{equation}
\mu(n+1)=[\mu(n)-\triangle_{\mu}(P_{1,\max}-P)]^{+}\label{eqn19}
\end{equation}
\begin{equation}
\vartheta(n+1)=[\vartheta(n)-\triangle_{\vartheta}(\tau_{\max}-\tau)]^{+}\label{eqn23}
\end{equation}
\begin{equation}
\nu(n+1)=[\nu(n)-\triangle_{\nu}(T-\tau)]^{+}\label{eqn20}
\end{equation}
and
\begin{equation}
\upsilon(n+1)=[\upsilon(n)-\triangle_{\upsilon}(\bar{R}(P,\tau)-r_{\min})]^{+}\label{eqn21}
\end{equation}
where $n$ is the iteration index, and $\triangle_{\mu}$,
$\triangle_{\nu}$ and $\triangle_{\upsilon}$ are the positive
iteration steps. Inspired by the Dinkelbach method
\cite{Dinkelbach}, we propose an iterative algorithm as follows
\rule{8.76cm}{1pt}\\
Algorithm 1: Energy-Efficient Resource Allocation\\
\rule{8.76cm}{1pt}
\begin{enumerate}

\item Initialization: Given $N_t$, $W$, $T$, $\eta$, $\alpha$,
$\theta$, $P_0$, $P_{1,\max}$, $P_{2,\max}$, $r_{\min}$,
$\triangle_{\mu}$, $\triangle_{\nu}$ and $\triangle_{\upsilon}$. Let
$\mu=0$, $\vartheta=0$, $\nu=0$, $\upsilon=0$, $P=0$ and
$q^{\star}=\frac{\bar{R}(P,\tau)T}{P_0T+P\tau}$. $\epsilon$ is a
sufficiently small positive real number.

\item Update $\mu$, $\vartheta$, $\nu$ and $\upsilon$ according to
(\ref{eqn19}), (\ref{eqn23}), (\ref{eqn20}) and (\ref{eqn21}),
respectively.

\item Computing the optimal $P^{\star}$ and $\tau^{\star}$ by
jointly solving the equations (\ref{eqn17}) and (\ref{eqn18}).

\item If
$\bar{R}(P^{\star},\tau^{\star})T-q^{\star}(P_0T+P^{\star}\tau^{\star})>\epsilon$,
then set
$q^{\star}=\frac{\bar{R}(P^{\star},\tau^{\star})T}{P_0T+P^{\star}\tau^{\star}}$,
and go to 2). Otherwise, $P^{\star}$ is the optimal transmit power
and $\tau^{\star}$ is the optimal transfer duration.
\end{enumerate}
\rule{8.76cm}{1pt}

\section{Numerical Results}
To examine the effectiveness of the proposed energy-efficient
resource allocation scheme, we present several numerical results in
the following scenarios: we set $W=10$ KHz, $T=5$ ms, $\eta=0.8$,
$\sigma^2=1$, $r_{\min}=12$ Kb/s, $P_0=45$Watt and
$P_{1,\max}=P_{2,\max}=15$Watt. In addition, we set $\alpha=\theta$
for convention. It is found that the proposed energy-efficient
resource allocation scheme converges after no more than 20 times
iterative computation in the all simulation scenarios.

\begin{figure}[h] \centering
\includegraphics [width=0.5\textwidth] {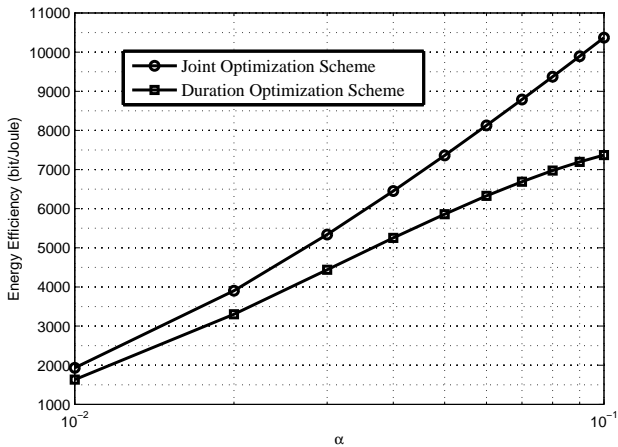}
\caption {Performance comparison of the proposed and the fixed
resource allocation schemes.} \label{Fig2}
\end{figure}

Fig.\ref{Fig2} compares the energy efficiencies of the proposed
power and duration joint optimization scheme and duration
optimization scheme with $N_t=100$. Intuitively, it is optimal to
use $P_{1,\max}$ as the transmit power at S$_1$ in the sense of
maximizing the harvesting energy, so we set $P=P_{1,\max}$ and
optimize $\tau$ only for the duration optimization scheme based on
Algorithm 1. As seen in Fig.\ref{Fig2}, the joint optimization
scheme performs better than the duration optimization one obviously,
as the latter fixes the transmit power, which also has a greater
impact on energy efficiency. For example, when $\alpha=0.05$, there
is about $1.5$Kb/J gain. As $\alpha$ increases, the performance gain
becomes larger. Therefore, the proposed scheme can effectively
increase the energy efficiency of wireless information and power
transfer.

\begin{figure}[h] \centering
\includegraphics [width=0.5\textwidth] {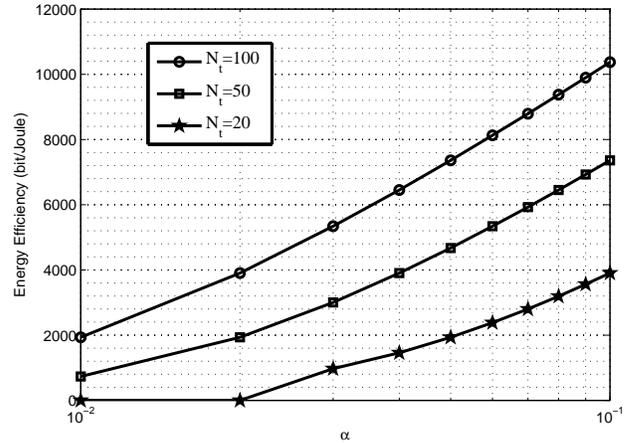}
\caption {Performance comparison of the proposed resource allocation
scheme with different numbers of BS antennas.} \label{Fig3}
\end{figure}

Fig.\ref{Fig3} investigates the effect of the number of antennas at
S$_1$ on the energy efficiency of the proposed scheme. Clearly, as
$N_t$ increases, there is significant performance gain in terms of
energy efficiency, especially for the large $\alpha$. For example,
when $\alpha=0.05$, there is about $2.5$Kb/J gain by increasing
$N_t$ from 50 to 100. It is found that when $\alpha\leq0.02$, the
energy efficiency in the case of $N_t=20$ reduces to zero. This is
because the harvested power can not support the given QoS
requirement under such an antenna setup. However, with the increase
of antenna number, i.e $N_t=50$, the wireless information and power
transfer system can support the QoS guaranteed service even with
$\alpha=0.01$. Hence, the proposed scheme can realize long-distance
and high QoS provisioning wireless information and power transfer.

\section{Conclusion}
A major contribution of this letter is the introduction of the
large-scale MIMO technique into wireless information and power
transfer. By exploiting the advantage of the large-scale MIMO, this
letter realizes long-distance and QoS guaranteed wireless
information and power transfer. Considering the demand for green
communications, an energy-efficient resource allocation scheme is
proposed by jointly optimizing transmit power and transfer duration.
Numerical results confirm the effectiveness of the proposed scheme.


\begin{thebibliography}{1}
\bibitem{Medical}
F. Zhang, S. A. Hackworth, X. Liu, H. Chen, R. J. Sclabassi, and M.
Sun, ``Wireless energy transfer platform for medical sensor and
implantable devices," in \emph{Proc. IEEE EMBS 31st Annual Int.
Conf.}, pp. 1045-1048, Sep. 2009.

\bibitem{WET0}
K. Huang, and E. G. Larsson, ``Simultaneous information and power
transfer for broadband wireless systems," [online]:
http://arxiv.org/pdf/1211.6868v2.pdf

\bibitem{WET1}
M. K. Watfa, H. Al-Hassanieh, S. Selman, ``Multi-hop wireless energy
transfer in WSNs," \emph{IEEE Commun. Lett.}, vol. 15, no. 12, pp.
1275-1277, Sep. 2011.

\bibitem{WET2}
L. R. Varshney, ``Transporting information and energy
simultaneously," in \emph{Proc. IEEE Int. Symp. Inf. Theory (ISIT)},
pp. 1612-1616, July 2008.

\bibitem{WET3}
J. Gozalvez, ``Witricity-the wireless power transfer," \emph{IEEE
Veh. Tech. Magazine}, vol. 2. no. 2, pp. 38-44, June 2007.

\bibitem{WET4}
P. Grover, and A. Sahai, ``Shannon meets Tesla: wireless information
and power transfer," in \emph{Proc. IEEE Int. Symp. Inf. Theory
(ISIT)}, pp. 2363-2367, June 2010.

\bibitem{EnergyBeamforming1}
R. Zhang, and C. K. Ho, ``MIMO broadcasting for simultaneous
wireless information and power transfer," in \emph{Proc. IEEE
Globecom}, pp. 1-5, Dec. 2011.

\bibitem{EnergyBeamforming2}
Z. Xiang, and M. Tao, ``Robust beamforming for wireless information
and power transmission," \emph{IEEE Wireless Commun. Lett.}, vol. 1,
no. 4, pp. 372-375, Aug. 2012.

\bibitem{EnergyEfficiency}
X. Chen, C. Yuen, and Z. Zhang, ``Wireless energy and information
transfer tradeoff for limited feedback multi-antenna systems with
energy beamforming," \emph{IEEE Trans. Veh. Tech.}, 2013.

\bibitem{LargescaleMIMO1}
T. L. Marzetta, ``Noncooperative cellular wireless with unlimited
number of base station antennas," \emph{IEEE Trans. Wireless
Commun.}, vol. 9, no. 11, pp. 3590-3600, Nov. 2010.

\bibitem{LargescaleMIMO2}
F. Rusek, D. Persson, B. K. Lau, E. G. Larsson, T. L. Marzetta, O.
Edfors, and F. Tufvesson, ``Scaling to MIMO: opportunities and
challenges with very large arrays," \emph{IEEE Signal Process.
Mag.}, vol. 20, no. 1, pp. 40-60, Jan. 2013.

\bibitem{Green}
F. Chu, K. Chen, and G. Fettweis, ``Green resource allocation to
minimize receiving energy in OFDMA cellular systems," \emph{IEEE
Commun. Lett.}, vol. 16, no. 3, pp. 372-374, Jan. 2012.

\bibitem{ChannelHardening}
B. M. Hochwald, T. L. Marzetta, and V. Tarokh, ``Multiple-antenna
channel hardening and its implications for rate-feedback and
scheduling," \emph{IEEE Tran. Inf. Theory}, vol. 50, no. 9, pp.
1893-1909, Sept. 2004.

\bibitem{QoS}
X. Chen, Z. Zhang, S. Chen, and C. Wang, ``Adaptive mode selection
for multiuser MIMO downlink employing rateless codes with QoS
provisioning," \emph{IEEE Trans. Wireless Commun.}, vol. 11, no.
2, pp. 790-799, Feb. 2012.

\bibitem{EnergyEfficientPLS}
D. W. K. Ng, E. S. Lo, and R. Schober, ``Energy-efficient resource
allocation for secure OFDMA systems," \emph{IEEE Trans. Veh.
Tech.}, vol. 61, no. 6, pp. 2572-2585, Jul. 2012.

\bibitem{Dinkelbach}
W. Dinkelbach, ``On nonlinear fractional programming," \emph{Manage.
Sci.}, vol. 13, no. 7, pp. 492¨C498, Mar. 1967. [Online]. Available:
http://www. jstor.org/stable/2627691


\end{thebibliography}
\end{document}